\documentclass[a4paper,10pt,prl,twocolumn,superscriptaddress,nofootinbib]{revtex4-1}

\usepackage[dvips]{graphicx}
\usepackage[dvips,usenames]{color}
\usepackage{amsmath}
\usepackage{amsfonts}
\usepackage{txfonts}
\usepackage{amssymb}
\usepackage{amstext}
\usepackage[sort&compress]{natbib}
\usepackage{graphics,graphicx}

\newcommand{\be}{\begin{equation}}
\newcommand{\ee}{\end{equation}}
\newcommand{\bea}{\begin{eqnarray}}
\newcommand{\eea}{\end{eqnarray}}
\newcommand{\ket}[1]{\left\vert #1    \right\rangle }

\newcommand{\curly}[1]{\left\{{#1}\right\}}
\newcommand{\crc}[1]{\left({#1}\right)}
\newcommand{\abs}[1]{\left\vert{#1}\right\vert}
\newcommand{\expectation}[1]{\left\langle #1   \right\rangle }

\def\etal{{\it et al.\/}}
\newcommand{\dg}{^{\dagger}}
\newcommand{\w}{\omega}
\newcommand{\W}{\Omega}

\newcommand{\Hbase}{H_{OM}}
\newcommand{\Hlinear}{H_{lin.}}

\newcommand{\PRLsection}[1]{\emph{#1} - }
\newcommand{\PRLsubsection}[1]{\emph{#1} - }

\begin{document}

\title{Pulsed Laser Cooling for Cavity-Optomechanical Resonators}
\author{S. Machnes}
\email{shai.machnes@uni-ulm.de}
\affiliation{Institut f{\"u}r Theoretische Physik, Universit{\"a}t
Ulm, D-89069 Ulm, Germany}
\author{J. Cerrillo}
\affiliation{QOLS,
The Blackett Laboratory, Imperial College London, Prince Consort
Rd., SW7 2BW, UK}
\author{M. Aspelmeyer}
\affiliation{Vienna Center for Quantum Science and Technology, Faculty of Physics, University of Vienna, Boltzmanngasse 5, A-1090 Vienna, Austria.}
\author{W. Wieczorek}
\affiliation{Vienna Center for Quantum Science and Technology, Faculty of Physics, University of Vienna, Boltzmanngasse 5, A-1090 Vienna, Austria.}
\author{M. B. Plenio}
\affiliation{Institut f{\"u}r Theoretische Physik, Universit{\"a}t
Ulm, D-89069 Ulm, Germany}
\affiliation{QOLS,
The Blackett Laboratory, Imperial College London, Prince Consort
Rd., SW7 2BW, UK}
\author{A. Retzker}
\affiliation{Institut f{\"u}r Theoretische Physik, Universit{\"a}t
Ulm, D-89069 Ulm, Germany}

\begin{abstract}
A pulsed cooling scheme for optomechanical systems is presented that is capable of cooling at much faster rates, shorter overall cooling times, and for a wider set of experimental scenarios than is possible by conventional methods. The proposed scheme can be implemented for both strongly and weakly coupled optomechanical systems in both weakly and highly dissipative cavities.
We study analytically its underlying working mechanism, which is based on interferometric control of optomechanical interactions, and we demonstrate its efficiency with pulse sequences that are obtained by using methods from optimal control. The short time in which our scheme approaches the optomechanical ground state allows for a significant relaxation of current experimental constraints. Finally, the framework presented here can be used to create a rich variety of optomechanical interactions and hence offers a novel, readily available toolbox for fast optomechanical quantum control.
\end{abstract}

\maketitle

\PRLsection{Introduction}
Micro- and nanomechanical resonators are currently emerging as new quantum systems~\cite{Cleland2010}. Their integrability in a solid state architecture offers attractive opportunities for quantum information objectives such as mechanical quantum registers~\cite{Geller2004, Rabl2010}, optomechanical quantum transducers~\cite{Stannigel2010} or quantum memories~\cite{Chang2011}. At the same time, their size and mass promise access to a hitherto untested regime of macroscopic quantum physics~\cite{bose1997, Marshall2003, Pikovski2012, Romero2011}. The field of cavity quantum optomechanics~\cite{Kippenberg2008, Marq2009, MGV, Aspelm2010} utilizes methods from quantum optics in combination with optomechanical radiation pressure interactions to achieve this and experiments are progressing rapidly - the strong coupling regime has been demonstrated~\cite{Groblacher, Teufel, Verhagen2012}, and optomechanical analogues of electromagnetically induced transparency~\cite{Kipp2010, Paintre2011} have demonstrated first steps towards mechanical storage of light.

A prerequisite to achieve full coherent control over mechanical quantum states is to operate these systems close to their quantum ground state and to achieve coupling rates that exceed all other decoherence rates. Optomechanical cooling close to ~\cite{schwab09,grobl09,kipp09,Paintre2011,Verhagen2012} and even well into ~\cite{Teufel2011,ChanNature2011} the quantum ground state of micro- and nano-mechanical devices have been realized. However, most mechanical devices are intrinsically connected to a hot environment through their supports, which results in large heating rates. Efficient cooling therefore requires to minimize the thermal coupling, either by operating in a cryogenic environment \cite{Groblacher08, schwab09, grobl09, kipp09, Teufel2011, ChanNature2011} or by decoupling the mechanical resonator from its environment \cite{Zoller, Cirac1, Barker}.

As a consequence, a cooling scheme that can beat the mechanical heating rate for systems that are not cryogenically cooled or mechanically decoupled from the environment is highly desirable. For the most widely used scheme, sideband cooling \cite{Wilson-Rae,Marquardt2008,Wineland,Genes2008}, the cooling rate $\Gamma$, and conversely the time required to approach the ground state, are inherently limited by the mechanical frequency, $\nu$. Specifically $\Gamma < \nu$, due to use of the rotating wave approximation (RWA). Recently it was demonstrated in the context of ion trap physics that pulsed schemes can break the speed limit set by the oscillator frequency \cite{SuperFast}, by using interference between optical pulses incident on the system that is being cooled. This generates an effective cooling (red-sideband) term with large couplings, thus avoiding the RWA limitations. Unfortunately, this approach cannot be directly used as the nature of the cavity coupling is significantly different (generally non-linear for the optomechanical system, and when linearized, h.o. (harmonic oscillator) to h.o. coupling v. h.o. to spin coupling for trapped ions) as is the nature of the subsystem from which energy is eventually removed (optical cavity with a wide range of quality factors, v.s. the ion with a finite Hilbert space and fixed physical properties). Moreover, one needs to overcome the instability issue in optomechanical systems, which does not exist for trapped ions.

In this letter we demonstrate how a sequence of fast pulses adds a term to the effective optomechanical interaction Hamiltonian which approximates the cooling (also known as beam-splitter, anti-Stokes and red sideband) operator $x_m x_c + p_m p_c \propto a b\dg + a\dg b$. Here $a$ is the annihilation operator of the cavity and $x_c$,$p_c$ are its quadrature operators, $b$ is the annihilation operator of the mechanical oscillator with corresponding quadratures $x_m$,$p_m$. The technique is shown to be experimentally feasible in both the good cavity ($\kappa\ll\nu$) limit, where it is capable of reaching the ground-state much faster than the oscillator frequency, and the bad cavity limit ($\kappa>\nu$), where sideband cooling is incapable of approaching the ground state.

\PRLsection{Physical Setting}
Cavity optomechanical systems are modeled as an optical cavity field which couples to a mechanical resonator by way of radiation pressure. The optical (mechanical) mode, oscillating at a frequency $\w$ ($\nu$)  is characterized by a relaxation rate $\kappa$ ($\gamma_{m}$). The optical mode is driven by a detuned laser field of frequency $\w_{l}$ with a strength $\W$. The corresponding Hamiltonian is
\be
{\Hbase}=\Delta a\dg a+\nu b\dg b+\frac{g_{0}}{\sqrt 2}a\dg a\left(b\dg+b\right)+\W\left(a\dg e^{-i\phi} +a e^{i\phi}\right)
\label{Hamil}
\ee
where $\phi$ is the initial driving phase, $\Delta=\omega-\omega_l$ is the cavity detuning w.r.t the laser frequency, and $g_{0}$ is the optomechanical coupling rate.

\PRLsection{Linear Approach}
For continuous driving, cooling is achieved by invoking the RWA. In order to derive a pulse sequence which generates cooling directly via the cooling operator introduced above, i.e. without RWA, we will need to analyze the dynamics of the system. Assuming, for now, weak coupling, $g_{0}\ll\nu$, one may avoid the complexity of the non-linear nature of the interaction by considering a linearized approximation of the system.

As this scheme makes use of rapidly changing driving of the cavity mode, the usual Hamiltonian linearization procedure \cite{Groblacher,Wilson-Rae} cannot be trivially applied to eq.\,(\ref{Hamil}). Rather, we move to a frame co-moving with the state of the cavity, by applying a time-dependent canonical transformation (see Appendix A for additional details). In this frame, $a$ is redefined as a small perturbation, allowing us to replace the quadratic coupling with a linear one. This results in the Hamiltonian:
\be
{\Hlinear}=\Delta a\dg a + \nu b\dg b + \left(G\left(t\right)a+G^{*}\left(t\right)a\dg\right)x_{m}+\left|G\left(t\right)\right|^{2}x_{m}
\label{linear}
\ee
with $G\left(t\right)=ig_{0}e^{-i\left(\Delta-i\kappa\right) t}\int_{0}^{t}\Omega\left(t^{\prime}\right)e^{i\left(\Delta-i\kappa\right) t^{\prime}}dt^{\prime}$.

The linear nature of the dynamics now allows us to re-phrase the dynamics as an equation-of-motion of the covariance matrix \cite{MasterEq}. We define a vector of quadrature operators $\mathbf{R}\equiv(x_{c},p_{c},x_{m},p_{m})^{t}$ and the covariance matrix as
 $\gamma_{i,j}\equiv2Re\crc{\expectation{R_{i}R_{j}}-\expectation{R_{i}}\expectation{R_{j}}}$. The corresponding equation of motion is $\frac{d}{dt}\gamma=M\gamma+\gamma\left(M^{T}\right)+\frac{\kappa}{2}P$
 with $M=S V-\frac{\kappa}{2}P$ where $V$ is the potential matrix ($\Hlinear=R^{T}VR$), $S$ the symplectic matrix and $P=diag\crc{1,1,0,0}$.

Denoting a series of $n$ control pulses (i.e. $n$ modulations of the cavity driving laser), $\curly{H_{c1},{\ldots},H_{cn}}$ of corresponding durations $\crc{t_1{\dots}t_n}$ (with free evolution implicit), and using the Baker-Campbell-Hausdorff  (BCH) \cite{BCH} equivalency, one may compute the equivalent control Hamiltonian $H_{c}$ from
$e^{\frac{-i}{\hbar}t_{n}\crc{H_0+H_{cn}}}{\cdots}e^{\frac{-i}{\hbar}t_{1}\crc{H_0+H_{c1}}}=e^{\frac{-i}{\hbar}\left(\sum{t_{k}}\right)\crc{H_0+H_{c}}}$, where $H_{0}$ as in eq.\,(\ref{linear}) for $G=0$.

The equivalent control Hamiltonian to the control sequence $\curly{-Gx_{c}x_{m},+Gp_{c}x_{m},-Gp_{c}x_{m},-Gx_{c}x_{m}}$ for $\crc{t_2,t_1,t_1,t_2}$, with $\Delta=\nu$ (to second order in the BCH series) is
\be
H_{c}=G\crc{t_1^2 \nu - 2t_2}x_{c}x_{m} - G t_{1}^2 \nu p_{c}p_{m}.
\label{CoolSeq1}
\ee
 Setting $t_2=t_{1}^{2}\nu$, the desired $-G t_{1}^2 \nu\crc{x_{c}x_{m}+p_{c}p_{m}}$ cooling operator is achieved. While higher-order elements in the BCH series contribute terms undesirable for cooling (e.g. squeezing), both analytical estimates and numerical studies show these disruptions to be well-contained for pulse durations shorter than the mechanical frequency. Moreover, as will be shown later, optimal control can further reduce undesired terms, improving upon sideband cooling w.r.t both the achievable final temperature and the required time. Finally, it is important to stress that the BCH operators are a two-edged sword - they create the cooling operators from commutation relations between the driving terms and the free Hamiltonian, but at the same time they transform the dissipative elements (Lindblad terms) in a similar fashion. As a result, stronger driving enhances both cooling and dissipation terms (modified by the BCH relations), resulting in sub-linear advantage of such an approach to further increase cooling rates.

\begin{figure}
\begin{center}
\includegraphics[width=\columnwidth]{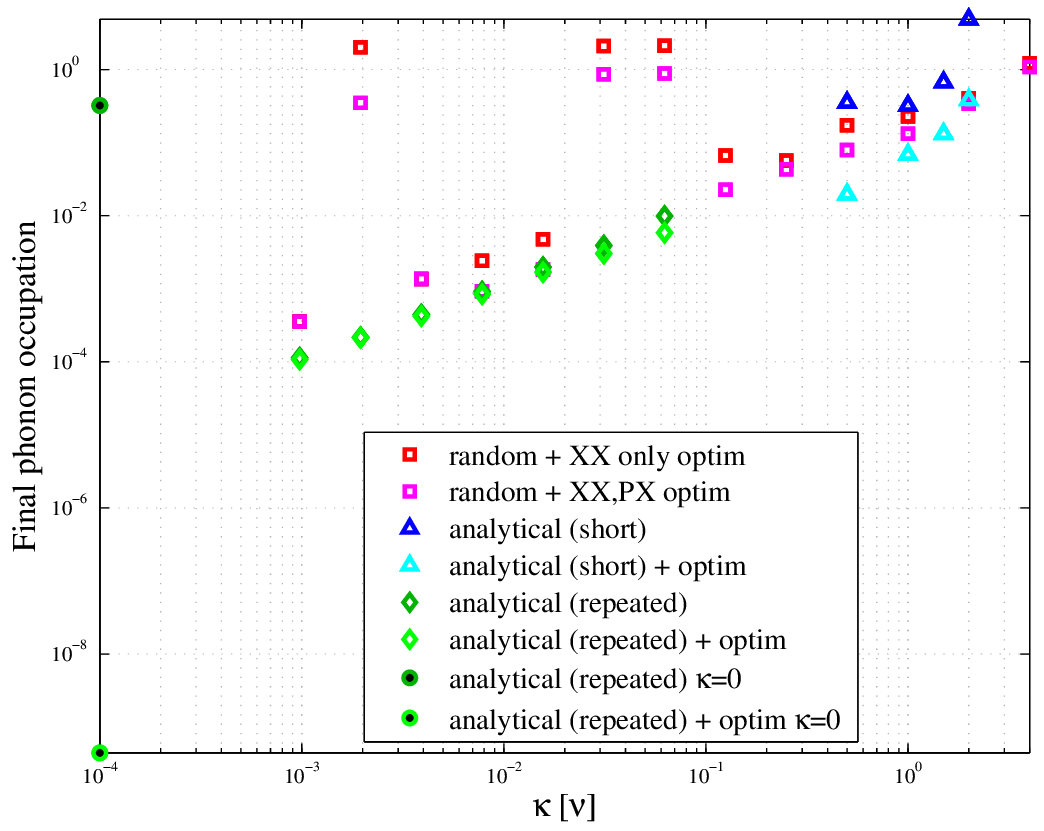}
\end{center}
\caption{
Results of several covariance matrix optimization procedures. All with an initial phonon thermal occupation of $10$, $\left|G\right|<10\nu$, $\gamma=0$ and taking at most $0.8\times\frac{2\pi}{\nu}$. The red and magenta data sets involve optimizations with random initial pulse-sequences, optimized with partial ($x_{c}x_{m}$-only) and full coupling respectively; the random initial settings and nature of global optimization is responsible for the poor performance at $\kappa\approx0.03$. Final temperatures achieved by sequences of 30 pulses (based on the four pulse analytical formulations, repeated 7.5 times) are shown before and after optimization (blue and cyan). Long cyclic sequences (75 repetitions of the 4 pulse analytical sequence) are shown before and after optimization and are represented by the dark and light green sets. Application of the last set, pre-optimization, for $\kappa=0$, appears at the left axis of the plot. Note that as often with numeric optimization, the points may represent local optima.}
\label{kappa}
\end{figure}

\begin{figure}
\begin{center}
\includegraphics[width=\columnwidth]{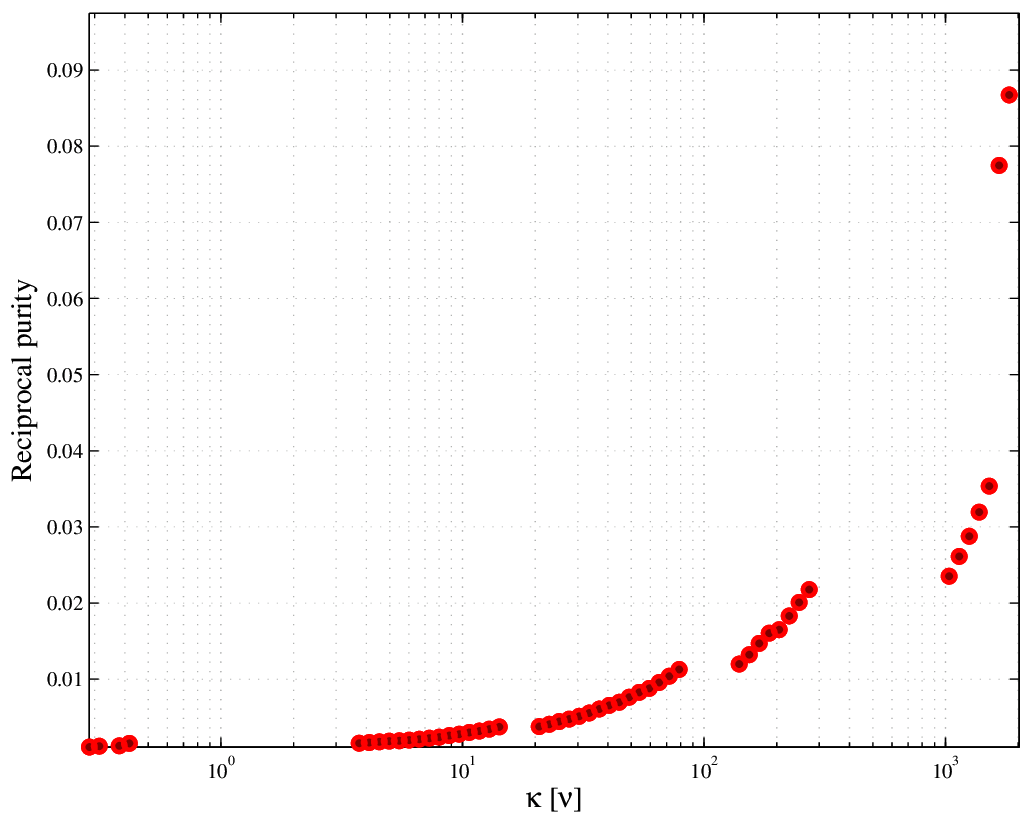}
\end{center}
\caption{
Performance of optimized cooling sequences for $\kappa>\nu$. Starting with the analytically derived sequences, a series of optimizations was used to generate cooling sequences for successively higher dissipative systems, with the result of one optimization serving as the starting point for the next. Note that in this case, the resulting system state is squeezed, as a result of higher-order BCH terms. 
}
\label{highkappa}
\end{figure}

\PRLsubsection{Optimal Control}
We have examined the performance and limits of the proposed cooling sequence using search methods from optimal control. We took multiple approaches, including (a) two stage optimization - initially optimizing the pulse amplitude and subsequently optimizing both amplitude and phase; (b) using the analytically derived sequence in eq.\,(\ref{CoolSeq1}) as an initial point of the optimization;Ê (c) in case of strongly dissipating cavities, a "telescoping" series of optimizations is used, slowly increasing dissipation, with the sequence resulting from optimization $k$ serving as the initial condition for optimization $k+1$ (d) a similar series of optimization is used to gradually shorten overall cooling times (e) random starting conditions and simultaneous optimization of all control parameters. Fig.\,(\ref{kappa})  shows the results of multiple optimizations for varying values of $\kappa$, and a comparison to the results of the analytical sequence before numerical optimization. Fig.\,(\ref{highkappa}) shows optimization results for the bad cavity regime. There, we allow the system state to become squeezed alongside the desired cooling and optimize for the squeezed phonon count (measured by the reciprocal purity \cite{Purity1,Purity2}).

An example of such an optimal sequence has been obtained using the full range of interactions, $\left(Re[G]x_c+Im[G]p_c\right)x_m$. Starting with an initial thermal phonon occupation of $100$, and for $\kappa=0$, we are able to achieve a final occupation below
$2\times10^{-7}$ in less than $0.6{2\pi}{\nu}^{-1}$ (see Appendix D). Another example, illustrated in figure (\ref{kappa2k}) presents the detailed behavior of a highly dissipative cavity $\kappa=2167\nu$. For such settings, it is much harder to suppress all undesirable terms in the higher orders of the BCH series. However, the system can be cooled in under $10^{-4}\frac{2\pi}{\nu}$ to less than one (squeezed) phonon, if, again, we allow the system state to become squeezed. All optimizations have been performed using QLib \cite{QLib}.

\PRLsection{Non-Linear Case}
For systems with large coupling $g_{0}$ \cite{Largeg0,ding2010} it is necessary to treat the full nonlinear interaction as presented in eq.\,(\ref{Hamil}). Unfortunately, in the case of the single-cavity system analyzed thus far, it does not appear to be possible to generate a cooling operator using a simple sequence of pulses analogous to that used for the linearized system (see Appendix B).

To overcome this limitation, a double cavity device, where the oscillator is coupled to two identical cavities (on the left and right of the oscillator), is proposed \cite{Paternostro2007,Vitali,2cavities}. This geometrical arrangement results in the coupling of the two cavities to the oscillator to be of opposite sign, which allows one to cancel the mechanical oscillator driving using opposing radiation pressure of the two cavities. The undriven system  Hamiltonian is
\be
H_{0}= \Delta(a_{1}\dg a_{1} + a_{2}\dg a_{2})+  \nu b\dg b+ \frac{g_{0}}{\sqrt{2}} ( a_{1}\dg a_{1} - a_{2}\dg a_{2})(b\dg+b),
\ee
where the subscript distinguishes the two cavities and $\Delta$ is the detuning of both cavities. We define the symmetric and antisymmetric modes, $a_a=\frac{1}{\sqrt{2}}(a_1+a_2)$, $a_s=\frac{1}{\sqrt{2}}(a_1-a_2)$ and the dimensionless quadrature operators $x=\frac{1}{\sqrt{2}}\crc{a^{\dg}+a}$, $p=\frac{i}{\sqrt{2}}\crc{a^{\dg}-a}$,  re-expressing the Hamiltonian as
\be
H_{0}= \Delta(a_{a}\dg a_{a} + a_{s}\dg a_{s})+  \nu b\dg b+ g_0 ( x_ax_s + p_ap_s)x_m.
\label{fullnonlinearhamiltonian}
\ee

Building the proposed cooling sequence in stages, let us first examine the terms generated by a $\{-\Omega x_a, 0, \Omega x_a\}$ sequence, $g_{0}\Omega t_1 p_s x_m+2\Delta\W t_{1} p_{a}$. By driving the cavity during the free evolution period with a $p_{a}$ pulse, the coefficient for $p_{a}$, above, can be controlled at will. By defining this additional pulse as $-(\alpha+2)\Delta\W t_{1}p_{a}$  the generated terms transform to:
\be
H_{NL1}=g_0 \Omega t_1 p_s x_m-\alpha\Delta\W t_{1}p_{a}.
\label{subsequence}
\ee

The cooling sequence for the double-cavity setup is
\be
\{\{-\Omega x_a, \beta p_{a}, \Omega x_a\},0,\{\Omega x_a, \beta p_{a}, -\Omega x_a\}\},
\label{fullsequence}
\ee
where the nested notation emphasizes the nesting of sequences used. Here $\beta \equiv  -(\alpha+2)\Delta\W t_{1}$
and the pulse durations are $((t_1, t_f, t_1),t^{\prime}_f,(t_1, t_f, t_1))$ with $t_2\equiv t_1+t_f+t_1$ and $t_{3}\equiv 2t_2+t^{\prime}_{f}$.

Assuming $\Omega t_1 \gg 1$, the free evolution implicit in $H_{NL1}$ of eq.\,(\ref{subsequence}) can be neglected; the inverted sequence $\{\Omega x_a, 0, -\Omega x_a\}$ together with the additional driving during the free evolution will yield eq.\,(\ref{subsequence}) with an inverted sign for $\Omega$, resulting in what can be viewed as a sequence-of-sequences, i.e. a nested pulse sequence. Defining $\tau_1\equiv \Omega t_1 t_2$, we get the terms generated by  eq.\,(\ref{fullsequence}) to be
\be\begin{split}
H_{NL2}= +(\alpha- 2) g_0\Delta\tau_1 x_s x_m+2 g_0\nu \tau_1 p_m p_s\\
+\nu g_0^{2}\tau_1^2 p_s^2 -3\Delta  g_0^{2} \tau_1^2 x_m^2- \tau_1 g_0^{2} x_a x_m^2\\
+2\alpha\tau_1 \Delta^{2}x_a.
\label{twocav}
\end{split}\ee
When $\Delta=2\nu/(\alpha-2)$ this Hamiltonian contains the cooling operator for the mechanical oscillator using the symmetric mode of the cavities, i.e. the $x_{s}x_{m}$ and $p_{s}p_{m}$ terms appear with identical coefficients. Setting $\alpha=4$ one may meet the resonance condition $\Delta=\nu$, which is important in some experimental layouts, as stronger interactions are achieved at resonance. An additional $x_{a}$ driving pulse, of area $-2\alpha\tau_1t_{3} \Delta^{2}$, can be added to the sequence to counter the last term in eq.\,(\ref{twocav}), provided the correction term and the entirety of the sequence in eq.\,(\ref{fullsequence}) are combined via the Trotter decomposition \cite{Trotter}, to suppress any high-order BCH terms. The two quadratic terms can be absorbed into the corresponding frequency terms by means of a Bogoliubov transformation.

To achieve ground-state cooling in a time shorter than the mechanical oscillation, following eq.\,(\ref{twocav}), we require the beam-splitter prefactor $\Omega g_0 \gg \textrm{max}\left(\nu,\kappa\right)^2$, choosing the $t_i$'s to all be one order of magnitude smaller than $1/\textrm{max}\left(\nu,\kappa\right)$, to suppress higher-order BCH terms. Finally, note that when $G\ngg\nu$ one may do without the periods of dedicated free evolution, although pre-factors will be different.

\begin{figure}
\begin{center}
\includegraphics[width=\columnwidth]{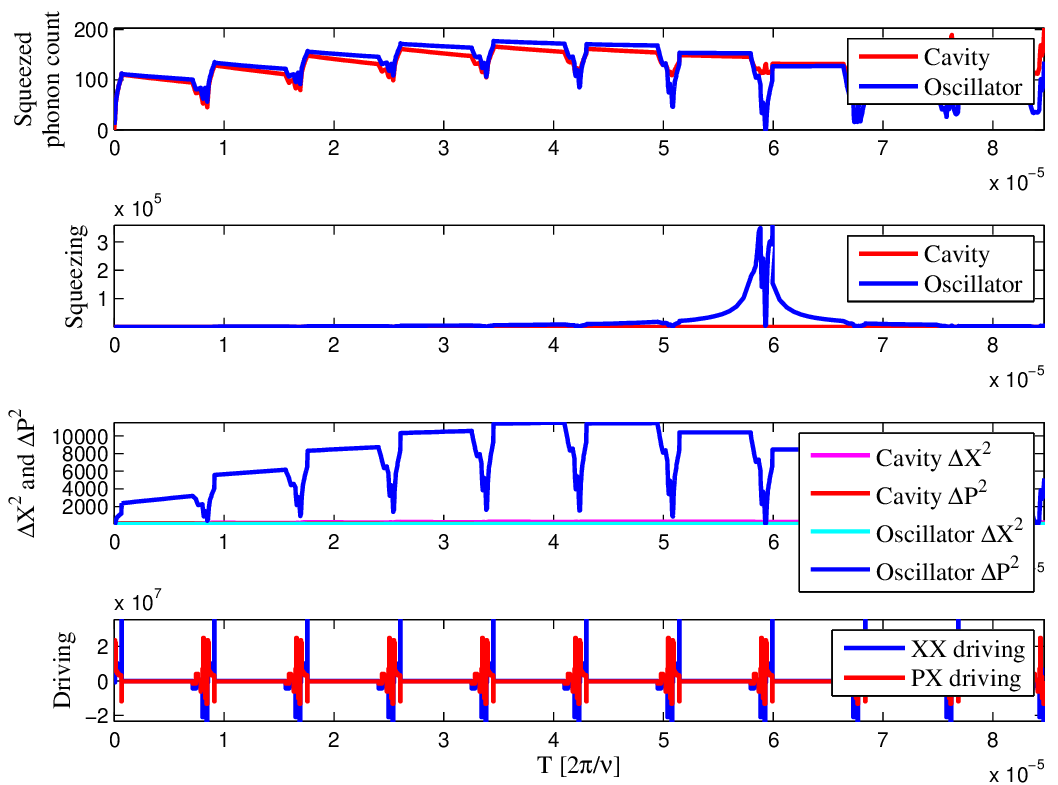}
\end{center}
\caption{Detailed behavior of highly optimized cooling pulses for  an optomechanical system with $\kappa=2167\nu$, with $10$ phonons in the initial state. Top panel contains the reciprocal purity for the mechanical oscillator and cavity; second panel the quadrature squeezing; third panel the covariance matrix elements and the forth panel details the driving. Note this is a regime in which sideband cooling is inappropriate, as the overall cooling time is orders of magnitude below $\nu$.}
\label{kappa2k}
\end{figure}

\PRLsection{Experimental feasibility}
To obtain both a cooling rate $\Gamma$ (defined as the pre-factor of the cooling operator in the Hamiltonian) beyond the limitation of continuous sideband-cooling, i.e. $\Gamma > \nu$, and to reach ground-state cooling faster than the mechanical frequency, readily available experimental parameters are sufficient: for an optomechanical Fabry-Perot cavity one can easily obtain $\nu=2\pi 10^6$ Hz, $m_{eff} = 5\times 10^{-11}$ kg and $\kappa=0.75\nu$~\cite{grobl09,Groblacher}, which yields $g_0=75$ Hz $\ll \nu$ and hence satisfies the linear regime of pulsed laser cooling (assuming a cavity length $L=10^{-2}$ m and an optical pump wavelength $\lambda=1064$ nm).

Previously achievable cooling rates have been of the order of $\Gamma\approx10^{-1}\times\nu$ \cite{grobl09}. With our method, and making use of optimal control, we fine-tune a 10-pulse cooling sequence of total duration $0.75\,\frac{2\pi}{\nu}$ and pulse energies of $\approx 40$ nJ per pulse, which can be created directly from amplitude modulating a $0.5 W$ continuous-wave laser beam, to obtain a cooling rate $\Gamma=1.3\,\nu$. For an initial temperature of $T=1K$ ($n_i\approx 2900$ phonons) and $\gamma_m\approx 300$Hz ($Q=2\times 10^4$) this sequence will reach mechanical thermal occupancies on the order of $n_f\approx 0.1$.

As a second example we consider a highly dissipative optomechanical Fabry-Perot double microcavity with $\nu=2\pi 10^4$ Hz, $m_{eff}=10^{-10}$ kg, $\kappa=2 \times 10^5 \, \nu$ and individual cavity lengths $L_1=L_2=4 \lambda$, as has been suggested in~\cite{VannerArXiv}. The resulting $g_0=10^6$ Hz $> \nu$ satisfies the nonlinear regime of pulsed laser cooling. Our method requires a set of laser pulses of length $\ll50\,$ps with a maximal peak power of $1$\,kW, which is available, for example, in the form of Q-switched lasers. This will result in a net cooling rate $\Gamma=10^4 \, \nu$. For $\gamma_m\approx 1$, i.e. a Q-factor of $6\times 10^4$, this would already allow cooling to the quantum ground state starting from room temperature.

\PRLsection{Conclusions}
We have introduced a novel pulsed cooling method for mechanical oscillators, which surpasses the intrinsic limits of conventional continuously pumped cooling. Our scheme is based on generating the cooling interaction, by quantum interference of successive pulses.  While already a simple analytical approach provides otherwise unachievable cooling rates $\Gamma > \nu$, the use of optimal control methods can further enhance these rates. We have also shown that current optomechanical configurations could achieve dramatic improvements in their experimental performance.
Following the methodology presented in this work, it is possible to generate a rich class of optomechanical interactions, for example the down-conversion (two-mode squeezing) interaction or various non-linear terms. This establishes a new and complete tool kit for fast preparation and manipulation of optomechanical quantum states and may very well provide a route towards room temperature quantum optomechanics.

\acknowledgments
This work was supported by the AXA Research Fund, by the EU STREP projects HIP, PICC, MINOS and EU project QESSENCE, by the Austrian Science Fund (FOQUS, START), by the European Research Council (ERC StG), as well as by the Alexander von Humboldt foundation.

\bigskip

While finishing this paper we learned of \cite{Wang}, which also treats pulsed cooling schemes for optomechanical systems.

\section{Appendix A - The linearization procedure in the strong-driving limit}

Starting with a simplified view of the cavity (limited to $X$ driving only)
\be
H= \Delta(a\dg a)+\W(t) (a+a\dg)
\ee
we derive the expectation value $\alpha$ of $a(t)$:
\be
\begin{split}
\dot a(t)&=i[H,a]=-i\Delta a-i\W\crc{t}\\
\dot\alpha&=-i\Delta \alpha-i\W\crc{t}\\
\alpha(t)&=\alpha_{0}e^{-i\Delta t} -ie^{-i\Delta t}\int_{0}^{t}\W\crc{t^\prime}e^{i\Delta t'}dt'\\
&\equiv\alpha_{0}e^{-i\Delta t} -f\crc{t}
\end{split}
\label{alpha}
\ee
We apply a displacement, $D=e^{\alpha^{*} a- \alpha a\dg}$, to a state-vector $\tilde{\ket{\Psi}}=D\ket\Psi$, and derive the transformed Hamiltonian $\tilde H$ by expanding the LHS of the Scr\"odinger equation for the displaced vector, $\frac{d}{dt}\tilde{\ket{\Psi}}=-i\tilde H\tilde{\ket{\Psi}}$. The result is $(\dot\alpha^{*}a-\dot\alpha a\dg) \tilde{\ket\Psi} -i DHD\dg \tilde{\ket\Psi}=-i\tilde H\tilde{\ket{\Psi}}
$, which implies

\be
\begin{split}
\tilde{H}&=DHD\dg+i\crc{\dot\alpha^{*}a-\dot\alpha a\dg} \\
&=\Delta\crc{a\dg a}+\abs{\alpha}^{2}\Delta+2\W\crc{t} \crc{\alpha+\alpha^{*}}
\end{split}
\ee

Turning our attention to the interaction term (which we ignored in the discussion above), one can show the coupling will be proportional to
\be
g\crc{a\dg a}+g\crc{\alpha a\dg+\alpha^{*}a}
\ee

Finally, note that the exercise above can be repeated for the dissipative system, replacing eq.\,(\ref{alpha}) with $\dot{\alpha}=-\kappa \alpha-i\Delta \alpha-i\W\crc{t}$ and deriving the modified Liouvillian $\tilde{L}$.

\section{Appendix B - Analytical approach to non-linear single-cavity cooling}

Let us analyze the effect of applying the pulse sequence appropriate to the linearized system to the non-linear case. As the $p_{c}x_{m}$ pulse is not directly available in the non-linear case, it must be generated by a BCH subsequence $\curly{-{\Omega}x_{c},H_0,{\Omega}x_{c}}$. Unfortunately, this sequence also introduces a mechanical driving term of the form $g_{0}\left(\Omega t_{p}\right)^{2}x_{m}$ that cannot be countered directly by cavity driving, but must be removed by displacing the mechanical operators.

Operators involving both the mechanical oscillator and the cavity present another issue, as they generate terms cubic in the position quadrature of the oscillator and render the system unstable. Specifically, when nesting BCH operations, i.e. applying for a time $t_{1}$ the pulse $2g_{0}\Omega t_{p} p_{c}x_{m}$ generated by the subsequence, in order to obtain the beam-splitter operator. The BCH relations imply the substitutions $x_{c}\rightarrow x_{c} -\Omega x_{m}$ and $p_{m}\rightarrow p_{m}+\Omega p_{c}$. Defining $\Omega'\equiv2g_{0}\Omega t_{p}$ the effective Hamiltonian then reads:
\be
\begin{split}
H=H_{0}+2\nu\Omega' t_{1}p_{c}p_{m}-2\Delta\Omega' t_{1} x_{c}x_{m}\\
+\nu \left(\Omega' t_{1}\right)^{2} p_{c}^{2}+\Delta \left(\Omega' t_{1}\right)^{2}x_{m}^{2}\\
-2g_0\Omega' t_{1}x_{c}x_{m}^{2}+g_0\left(\Omega' t_{1}\right)^{2}x_{m}^{3}.
\end{split}
\ee
The terms in the first line correspond to the beam-splitter operator when $\Delta=\nu$. The second line presents quadratic terms, for which a Bogoliubov transformation would be in order to absorb them to the frequency terms of the original Hamiltonian. The cubic terms in the last line are proportional to $g_{0}$. In the case of systems with $g_{0}\ll \nu$, these terms can be neglected and hence the linear approximation survives.

\section{Appendix C - Cavity dynamics}

A differential equation for the amplitude of the field in the cavity $E_{in}\left(t\right)$ as a function of a variable external field $E_{out}\left(t\right)$ of frequency $\omega$ is derived here. First of all, a recursion formula can be found by considering the state of the field inside the cavity after a round trip time $\tau\equiv\frac{2L}{c}$, with $L$ the length of the cavity. It can be considered to be the sum of (a) the previous field after bouncing off the mirrors of each end of the cavity (of reflectivity coefficient $\sqrt R$) and having picked up the phase $e^{i\omega\tau}$ and (b) the field that has entered the cavity during this time through the front quarter-wavelength mirror of transmissivity $\sqrt T$. This can be expressed as:
\be
E_{in}\left(t+\tau\right)=Re^{i\omega\tau}E_{in}\left(t\right)+i\sqrt{T}E_{out}\left(t+\tau\right).
\ee

The detuning $\Delta=\omega-\omega_{c}$, with $\omega_{c}=\frac{\pi}{L}c$, provides us with the only relevant phase difference:
\be
E_{in}\left(t+\tau\right)=Re^{i\Delta\tau}E_{in}\left(t\right)+i\sqrt{T}E_{out}\left(t+\tau\right).
\ee
Assuming that the driving laser is almost in resonance with the cavity eigenfrequency ($\Delta\tau\ll1$) the phase term can be Taylor expanded to first order:
\be
E_{in}\left(t+\tau\right)=R\left(1+i\Delta\tau\right)E_{in}\left(t\right)+i\sqrt{T}E_{out}\left(t+\tau\right).
\ee
For lossless cavities $R=1-T\approx1$, and therefore $T\Delta\tau$ terms can be neglected:
\be
\begin{split}
E_{in}\left(t+\tau\right)=E_{in}&\left(t\right)-TE_{in}\left(t\right)\\&+i\Delta\tau E_{in}\left(t\right)+i\sqrt{T}E_{out}\left(t+\tau\right).
\end{split}
\ee

In the limit where the dynamics in the cavity are negligible during the short $\tau$ timescale, a differential equation can be derived:
\be
\begin{split}
\frac{E_{in}\left(t+\tau\right)-E_{in}\left(t\right)}{\tau}&=-\frac{T}{\tau}E_{in}\left(t\right)\\+&i\Delta E_{in}\left(t\right)+i\frac{\sqrt{T}}{\tau}E_{out}\left(t+\tau\right).
\end{split}
\ee
Since $\kappa=\frac{T}{\tau}$:
\be
\frac{dE_{in}}{dt}=-\kappa E_{in}\left(t\right)+i\Delta E_{in}\left(t\right)+i\sqrt{\kappa}\frac{1}{\sqrt{\tau}}E_{out}.
\label{diff}
\ee

It is possible to relate this equation to the regular definition for the driving:
\be
\Omega=\sqrt{\frac{\kappa P}{\hbar\omega}}.
\ee
The power $P$ of an electromagnetic wave can be further expressed in terms of the field amplitude $E$ as $P=\frac{Energy}{time}=\frac{\epsilon_{0}E^{2}Acdt}{dt}=\epsilon_{0}E^{2}Ac$, where $A$ is the area of the cross section of the laser and $c$ the speed of light. Thus:
\be
\Omega=\sqrt{\frac{\kappa P}{\hbar\omega}}=\sqrt{\frac{\kappa\epsilon_{0}E_{out}^{2}A2L}{\hbar\omega\tau}}=\sqrt{\frac{\epsilon_{0}V}{\hbar\omega}}\sqrt{\frac{\kappa}{\tau}}E_{out}.
\ee
The pre-factor $\sqrt{\frac{\epsilon_{0}V}{\hbar\omega}}$ can be expressed as $\sqrt{\frac{\epsilon_{0}E^{2}V}{\hbar\omega}}\frac{1}{E}=\frac{\sqrt{n}}{E}$, so that it can be interpreted as the inverse of the electric field associated with one photon. The eq.(\ref{diff}) can be rewritten as:
\be
\begin{split}
\sqrt{\frac{\hbar\omega}{\epsilon_{0}V}}\frac{dE_{in}}{dt}=-\kappa\sqrt{\frac{\hbar\omega}{\epsilon_{0}V}}&E_{in}\left(t\right)\\&+i\Delta\sqrt{\frac{\hbar\omega}{\epsilon_{0}V}}E_{in}\left(t\right)+i\Omega
\end{split}
\ee
which is by definition:
\be
\frac{da}{dt}=-\kappa a+i\Delta a+i\Omega
\ee
with $a$ the field amplitude inside the cavity.

An alternative route to obtain the same result is to transform the square of the last term in eq.(\ref{diff}) into:
\be
\frac{E_{out}^{2}}{\tau}=\frac{n\left(t\right)}{\tau}\frac{E_{out}^{2}}{n\left(t\right)}=\frac{U}{\tau}\frac{1}{\hbar\omega}\frac{E_{out}^{2}}{n\left(t\right)}=\frac{P}{\hbar\omega}\frac{E_{out}^{2}}{n\left(t\right)}.
\ee
So the equation has to be rewritten as:
\be
\frac{dE_{in}}{dt}=-\kappa E_{in}+i\Delta E_{in}+i\sqrt{\frac{\kappa P}{\hbar\omega}}\frac{E_{out}}{\sqrt{n}}.
\ee
In order to get an equation for the field amplitudes, one has to divide by the field amplitude carried by one phonon ($\frac{E}{\sqrt{n}}$):
\be
\frac{da}{dt}=-\kappa a+i\Delta a+i\sqrt{\frac{\kappa P}{\hbar\omega}}.
\ee

\section{Appendix D - Sample optimized pulse sequence}
\begin{figure}
\begin{center}
\includegraphics[width=\columnwidth]{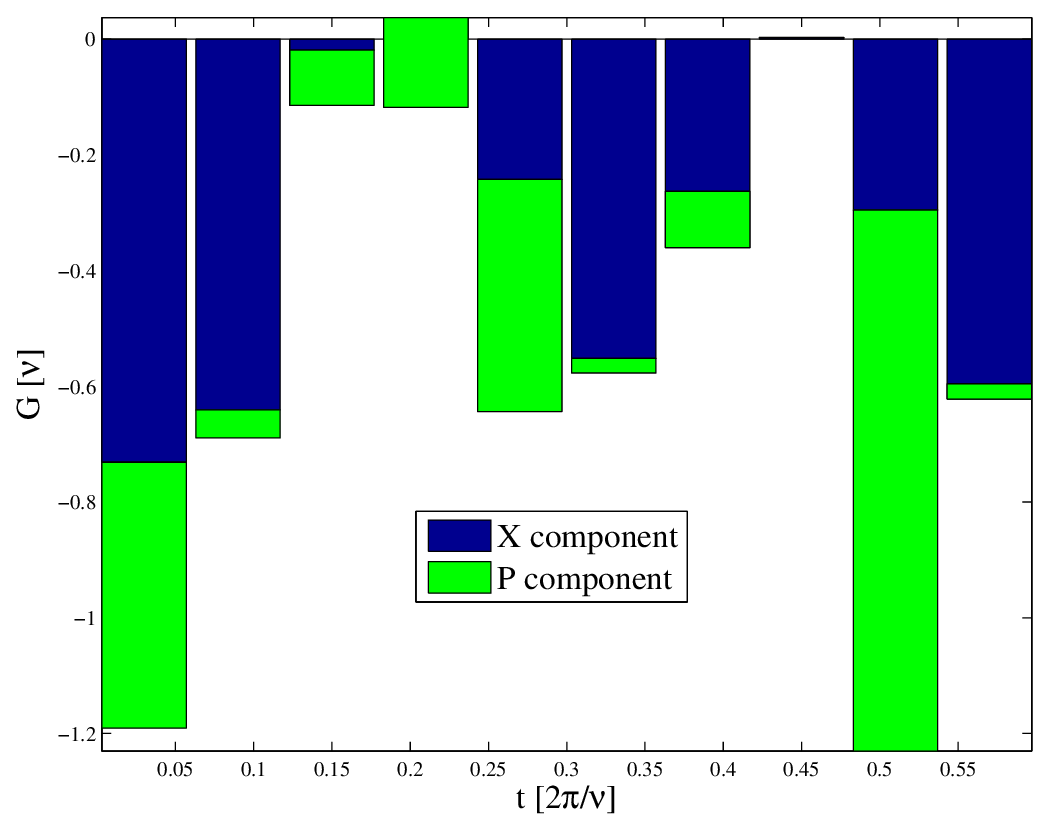}
\end{center}
\caption{Sample pulse sequence optimized for the full linear interaction, with initial phonon occupation of $100$, $G_{max}=\nu$, achieving a final occupation below $2\times10^{-7}$.}
\label{optseq}
\end{figure}

See fig. (\ref{optseq}).

\end{document}